\documentclass[11pt]{article}
\usepackage{amssymb,cite}
\newcommand{\be}{\begin{equation}}
\newcommand{\ee}{\end{equation}}
\newcommand{\bea}{\begin{eqnarray}}
\newcommand{\eea}{\end{eqnarray}}

\begin{document}

\title{Extension of Moyal-deformed hierarchies of soliton equations}

\author{Aristophanes Dimakis\, \footnote{Email: dimakis@aegean.gr} \\
 Department of Financial and Management Engineering, \\
 University of the Aegean, 31 Fostini Str., GR-82100 Chios, Greece
              \and
 Folkert M\"uller-Hoissen\, \footnote{Email: fmuelle@gwdg.de} \\
 Max-Planck-Institut f\"ur Str\"omungsforschung \\
 Bunsenstrasse 10, D-37073 G\"ottingen, Germany }

\date{ }

\maketitle

\begin{abstract}
Moyal-deformed hierarchies of soliton equations can be extended to larger
hierarchies by including additional evolution equations with respect to the
deformation parameters. A general framework is presented in which the extension
is universally determined and which applies to several deformed hierarchies,
including the noncommutative KP, modified KP, and Toda lattice hierarchy.
We prove a Birkhoff factorization relation for the extended ncKP and ncmKP
hierarchies. Also reductions of the latter hierarchies are briefly discussed.
Furthermore, some results concerning the extended ncKP hierarchy
are recalled from previous work.
\end{abstract}

\renewcommand{\theequation} {\arabic{section}.\arabic{equation}}

\section{Introduction}
\setcounter{equation}{0}
Classical soliton equations arise as compatibility conditions of linear systems
(see \cite{Blas98,Dick03}, for example).
Generalizing the ordinary product of functions of two coordinates $x,t$ to some
associative (and not necessarily commutative) product $\ast$, for which the partial
derivatives $\partial := \partial_x$ and $\partial_t$ are derivations, let us take
the linear system to be of the form
\be
    \psi_x = U \ast \psi \, , \qquad
    \psi_t = V \ast \psi
\ee
where $U,V$ are matrices of functions or suitable operators, depending
on some (matrix-valued) field $u(x,t)$.
The compatibility condition then reads
\be
    U_t - V_x + [ U , V ]_\ast = 0
\ee
where an index $x$ or $t$ indicates a corresponding partial derivative, and
$[U,V]_\ast := U \ast V - V \ast U$. If this equation is equivalent to a nonlinear
partial differential equation (PDE) for the dependent variable $u$, the differential
equation is said to admit a zero curvature formulation. If the latter is non-trivial,
the PDE possesses certain properties typically associated with completely integrable
models, in particular soliton equations. Generalizations of the underlying product
of functions in the aforementioned sense have been treated in particular in
\cite{Dorf+Foka92,Olve98,Kupe00,Stra01}. What happens if the product depends on the
coordinates or additional parameters?
An interesting example is given by the Moyal-product:
\be
   f \ast h := \mathbf{m}( e^{P/2} f \otimes h ) \, , \qquad
   P := \theta \, ( \partial_t \otimes \partial_x - \partial_x \otimes \partial_t )
   \label{Moyal}
\ee
where $\theta$ is a parameter, $\mathbf{m}(f \otimes h) = f h$ with the ordinary
product of the two functions $f,h$ on the right hand side, and the tensor product
has to be understood over $\mathbb{C}[[\theta]]$ (or with $\mathbb{C}$ replaced
by another field). The new parameter $\theta$ may now be treated on an equal footing
with the parameters $t$ and $x$. It is then natural to ask for an extension of the
above linear system by an additional equation of the form
\be
    \psi_\theta = W \ast \psi  \; .   \label{psi_theta}
\ee
Of course, this leads to additional compatibility conditions and a $\theta$-evolution
equation for the field $u$. In the examples we have explored so far \cite{DMH04hier,DMH04ncKP},
such an extension indeed exists, without restrictions imposed on the original linear system.
This new structure will be the central object in the following.
More precisely, we will concentrate on hierarchies, which arise from a linear system
of the above form and consist of evolution equations with `evolution times'
$t_n$, $n \in \mathbb{N}$, the flows of which commute with each other
(see section~\ref{sec:ext-MLh}).
In this case, a deformation parameter $\theta_{m,n}$ can be associated with each
pair of parameters $t_m,t_n$. The above procedure with $P$ in (\ref{Moyal})
replaced by
\be
   P := \sum_{m,n =1}^\infty \theta_{m,n} \, \partial_{t_m} \otimes \partial_{t_n}
        \label{Moyal2}
\ee
then results in an extension of this hierarchy by additional evolution
equations with respect to the parameters $\theta_{m,n} = - \theta_{n,m}$.
The new $\theta$-evolution equations are non-autonomous
equations.\footnote{Note that the classical KP hierarchy admits so-called `additional symmetries'
which are also non-autonomous since they depend explicitly on the variables $t_n$. See
\cite{Dick03}, for example.}
Fixing one equation of the extended hierarchy, all others are \emph{symmetries} of this
equation. The $\theta$-evolution equations are thus new symmetries of the Moyal-deformed
hierarchy equations.

Section~\ref{sec:ext-MLh} formulates an abstract scheme for the extension of
Moyal-deformed Lax hierarchies and so-called $N$-reductions (see \cite{Dick03},
for example) of the latter.
Section~\ref{sec:Sato} presents examples in the framework of Sato theory: the
extensions of the Moyal-deformed (matrix) KP \cite{DMH04hier,DMH04ncKP} and mKP
hierarchies, for which we also prove a Birkhoff factorization relation.
Another example is treated in section~\ref{sec:xncToda}: the extension of the
Moyal-deformed Toda lattice hierarchy.
Some remarks are collected in section~\ref{sec:remarks}.

\section{Extension of Moyal-deformed Lax hierarchies}
\setcounter{equation}{0}
\label{sec:ext-MLh}
Let $({\cal A}, \ast)$ be an associative algebra and ${\cal A} = {\cal A}_+ \oplus {\cal A}_-$
a decomposition of the Lie algebra $({\cal A}, [ \, , \, ]_\ast)$, where $[ \; , \; ]_\ast$ is
the commutator in the algebra $({\cal A},\ast)$, into Lie subalgebras ${\cal A}_{\pm}$, so that
\be
    \lbrack {\cal A}_- , {\cal A}_- \rbrack_\ast \subset {\cal A}_- \, , \qquad
    \lbrack {\cal A}_+ , {\cal A}_+ \rbrack_\ast \subset {\cal A}_+ \; .
      \label{decomp1}
\ee
Let $( \; )_+$ and $( \; )_-$ denote the projections to ${\cal A}_+$,
respectively ${\cal A}_-$.

For $L \in {\cal A}$, depending on functions of variables $t_n$, $n \in \mathbb{N}$,
consider the set of equations
\be
    L_{t_n} = [ (L^n)_+ , L ]_\ast
         \equiv - [ (L^n)_- , L ]_\ast
      \qquad n = 1,2, \ldots   \label{Lax-hier}
\ee
where $L^n = L \ast L \ast \ldots \ast L$. This implies
\be
   (L^m)_{t_n} = [ (L^n)_+ , L^m ]_\ast  \qquad m = 1,2, \ldots \; .  \label{L^m_tn}
\ee
Using the identity
\be
    [(L^n)_+ , (L^m)_-]_\ast + [(L^n)_- , (L^m)_+]_\ast
  + [(L^n)_+ , (L^m)_+]_\ast + [(L^n)_- , (L^m)_-]_\ast = 0 \qquad  \label{L-identity}
\ee
(which is obtained by decomposition of $[L^n,L^m]_\ast = 0$), (\ref{L^m_tn}),
and (\ref{decomp1}), we obtain
\be
   ((L^m)_+)_{t_n} - ((L^n)_+)_{t_m} + [(L^m)_+ , (L^n)_+]_\ast = 0  \label{+0curv}
\ee
with the help of which the commutativity of the flows (\ref{Lax-hier}) follows:
\bea
    (L_{t_n})_{t_m} - (L_{t_m})_{t_n}
  &=& [ (L^n)_+ , [ (L^m)_+ ,L ]_\ast ]_\ast + [ L , [ (L^n)_+ , (L^m)_+ ]_\ast ]_\ast
            \nonumber \\
  & &  + [ (L^m)_+ , [ L , (L^n)_+ ]_\ast ]_\ast
\eea
which vanishes due to the Jacobi identity.
The equations (\ref{Lax-hier}) and (\ref{+0curv}) are the integrability conditions
of the linear system
\be
    L \ast \psi = \lambda \, \psi \, , \qquad
    \psi_{t_n} = (L^n)_+ \ast \psi  \qquad n = 1,2,\ldots
\ee
where $\lambda_{t_n} = 0$ and $\psi$ lies in a domain on which the algebra $\cal A$
acts. So far, everything is well-known.
\vskip.2cm

Let us now specify the multiplication in $\cal A$ as the Moyal product (\ref{Moyal})
with (\ref{Moyal2}), and replace (\ref{decomp1}) by the stronger condition
\be
    {\cal A}_- \ast {\cal A}_- \subset {\cal A}_- \, , \qquad
    {\cal A}_+ \ast {\cal A}_+ \subset {\cal A}_+    \label{decomp}
\ee
so that ${\cal A}_-$ and ${\cal A}_+$ are \emph{subalgebras} of $({\cal A},\ast)$.
We extend the above linear system by appending linear equations of the form
\be
   \psi_{\theta_{m,n}} = W^{m,n} \ast \psi \; .   \label{ext-linsys}
\ee
This gives rise to additional integrability conditions, in particular
\be
   L_{\theta_{m,n}}
 = [W^{m,n},L]_\ast
   + {1 \over 2} \big( L_{t_n} \ast (L^m)_+ - L_{t_m} \ast (L^n)_+ \big)
    \label{L_theta_mn}
\ee
where we assumed $\lambda_{\theta_{m,n}} = 0$ and made use of the identity
\be
   (f \ast g)_{\theta_{m,n}} = f_{\theta_{m,n}} \ast g + f \ast g_{\theta_{m,n}}
   + {1 \over 2} ( f_{t_m} \ast g_{t_n} - f_{t_n} \ast g_{t_m} )
    \label{dtheta_star}
\ee
for functions $f,g$. For later use, we note that (\ref{L_theta_mn}) implies
\bea
   (L^k)_{\theta_{m,n}}
 = [ W^{m,n},L^k ]_\ast
   + {1 \over 2} \big( (L^k)_{t_n} \ast (L^m)_+ - (L^k)_{t_m} \ast (L^n)_+ \big)
   \label{L^k_theta_mn}
\eea
(see also \cite{DMH04hier}). The remaining integrability conditions
\bea
 0 &=& (W^{m,n})_{t_k} + [ W^{m,n} , (L^k)_+ ]_\ast - ((L^k)_+)_{\theta_{m,n}} \nonumber \\
   & & - \frac{1}{2} \big( \, ((L^k)_+)_{t_m} \ast (L^n)_+ - ((L^k)_+)_{t_n} \ast (L^m)_+
       \, \big)      \label{integr-cond1} \\
 0 &=& (W^{m,n})_{\theta_{r,s}} - (W^{r,s})_{\theta_{m,n}}
       + [ W^{m,n} , W^{r,s} ]_\ast
       + \frac{1}{2} \big( \, (W^{m,n})_{t_r} \ast (L^s)_+
       \qquad  \nonumber \\
   & & - (W^{m,n})_{t_s} \ast (L^r)_+
       - (W^{r,s})_{t_m} \ast (L^n)_+
       + (W^{r,s})_{t_n} \ast (L^m)_+ \, \big)  \label{integr-cond2}
\eea
require the commutativity of all the new flows (\ref{L_theta_mn}) and their commutativity
with the flows (\ref{Lax-hier}). If these equations can be satisfied with a suitable choice
of the operators $W^{m,n}$, then equations (\ref{L_theta_mn}) extend the hierarchy
(\ref{Lax-hier}) to a larger hierarchy of mutually commuting flows.
\vskip.2cm

For the moment, let us make the further assumption that
\be
      (L_{t_n})_+ = 0 = (L_{\theta_{m,n}})_+     \label{ass1}
\ee
i.e., $L_{t_n}, L_{\theta_{m,n}} \in {\cal A}_-$.
Then (\ref{Lax-hier}) implies $([ (L^n)_+ , L ]_\ast)_+ = 0$ and, using (\ref{decomp}),
(\ref{L_theta_mn}) leads to
\bea
 \Big( [W^{m,n},L]_\ast \Big)_+ &=& {1\over 2} \Big( L_{t_m} \ast (L^n)_+
     - L_{t_n} \ast (L^m)_+ \Big)_+ \nonumber \\
 &=& {1\over 2} \Big( [(L^m)_+ , L]_\ast \ast L^n
     - [(L^n)_+ , L]_\ast \ast L^m \Big)_+  \nonumber \\
 &=& {1\over 2} \Big( [(L^m)_+ \ast L^n
     - (L^n)_+ \ast L^m , L]_\ast \Big)_+  \nonumber \\
 &=& {1\over 2} \Big( [(L^n)_- \ast L^m - (L^m)_- \ast L^n , L]_\ast \Big)_+
\eea
so that
\be
 \Big( \Big[ W^{m,n} - {1\over 2} \big( (L^n)_- \ast L^m
   - (L^m)_- \ast L^n \big) , L \Big]_\ast \Big)_+ = 0 \; .
\ee
Assuming moreover
\be
     [L , {\cal A}_- ]_\ast \subset {\cal A}_-    \label{ass2}
\ee
(in accordance with (\ref{Lax-hier}) and (\ref{ass1})), we may take
$W^{m,n} \in {\cal A}_+$. The last equation is then satisfied if
\bea
  W^{m,n} &=& {1\over 2} \big( (L^n)_- \ast L^m - (L^m)_- \ast L^n \big)_+ \nonumber \\
          &=& {1\over 2} \big( (L^n)_- \ast (L^m)_+ - (L^m)_- \ast (L^n)_+ \big)_+ \; .
                    \label{Wmn}
\eea
It turns out that, with this choice, the flows (\ref{Lax-hier}) and (\ref{L_theta_mn})
indeed commute, so that the integrability conditions of the extended linear system
are satisfied.
\vskip.2cm

\noindent
{\bf Theorem.} Let ${\cal A} = {\cal A}_+ \oplus {\cal A}_-$ with subalgebras
${\cal A}_{\pm}$, and let $W^{m,n}$ be given by (\ref{Wmn}).
The equations (\ref{integr-cond1}) and (\ref{integr-cond2}) are
then consequences of (\ref{Lax-hier}) and (\ref{L_theta_mn}).
\vskip.1cm
\noindent
{\em Proof:} Using (\ref{L^k_theta_mn}), (\ref{L^m_tn}), (\ref{Wmn}) and (\ref{decomp}),
a straightforward calculation leads to
\bea
 & & ((L^k)_+)_{\theta_{m,n}} - [ W^{m,n} , (L^k)_+ ]_\ast
     + \frac{1}{2} \big( \, ((L^k)_+)_{t_m} \ast (L^n)_+ - ((L^k)_+)_{t_n} \ast (L^m)_+
       \, \big) \nonumber \\
 &=& \frac{1}{2} \, \big( \,
       ( [ (L^n)_- , (L^k)_-]_\ast + [ (L^n)_+ , (L^k)_-]_\ast )_- \ast (L^m)_+ \nonumber \\
 & & - ( [ (L^m)_- , (L^k)_-]_\ast + [ (L^m)_+ , (L^k)_-]_\ast )_- \ast (L^n)_+ \nonumber \\
 & & + (L^n)_- \ast [ (L^m)_+ , (L^k)_- ]_\ast
     - (L^m)_- \ast [ (L^n)_+ , (L^k)_- ]_\ast \, \big)_+ \; .   \nonumber
\eea
By use of the identity (\ref{L-identity}) this equals
\bea
   (W^{m,n})_{t_k} &=& \big( \, [ (L^m)_- , (L^k)_+ ]_\ast \ast (L^n)_+
     - [ (L^n)_- , (L^k)_+ ]_\ast \ast (L^m)_+  \nonumber \\
   & &  + (L^m)_- \ast [ L^n , (L^k)_+ ]_\ast
     - (L^n)_- \ast [ L^m , (L^k)_+ ]_\ast \, \big)_+   \nonumber
\eea
so that (\ref{integr-cond1}) holds. A similar, but tedious calculation, shows
that (\ref{integr-cond2}) is also satisfied.
\hfill $\blacksquare$
\vskip.2cm

Note that our intermediate assumptions (\ref{ass1}) and (\ref{ass2}) were not
needed in obtaining this result. Of course, only with a suitable choice of
$\cal A$ and $L$, the construction in this section will lead to meaningful
results. One has to ensure that (\ref{Lax-hier}) and (\ref{L_theta_mn}) are
sufficiently free of constraints.
Examples will be presented in the following sections.

\subsection{$N$-reductions}
\label{sec:Nred}
Using (\ref{L^m_tn}), (\ref{L^k_theta_mn}), and (\ref{decomp}), we obtain
\bea
     ( (L^N)_-)_{t_n}
 &=& ((L^N)_{t_n})_-
  = ( [ (L^n)_+ , L^N ]_\ast )_- \\
     ( (L^N)_-)_{\theta_{m,n}}
 &=& ( (L^N)_{\theta_{m,n}} )_- \nonumber \\
 &=& \frac{1}{2} \big( [ (L^n)_+ , (L^N)_- ]_\ast \ast (L^m)_+
     - [ (L^m)_+ , (L^N)_- ]_\ast \ast (L^n)_+ \big)_- \nonumber \\
 & & + ( [W^{m,n} , (L^N)_- ]_\ast )_-
\eea
which shows that the extended hierarchy preserves the \emph{$N$-reduction} constraint
(see \cite{Dick03}, for example)
\be
    (L^N)_- = 0  \qquad N \in \mathbb{N}  \; .
\ee
This leads to
\bea
  & & L_{t_{kN}} = 0 \, , \qquad
      L_{\theta_{kN,lN}} = 0  \qquad  \forall k,l \in \mathbb{N} \nonumber \\
  & & L_{\theta_{kN,lN+r}} = \frac{1}{2} \, L_{t_{(k+l)N+r}}
      \qquad \forall k \in \mathbb{N}, \; l \in \mathbb{N} \cup \{ 0 \},
      \; r = 1,2,\ldots,N-1  \, . \qquad
      \label{red-cond}
\eea
In particular, after reduction, on the variables $t_{(k+l)N+r}$ and $\theta_{kN,lN+r}$
the fields only depend through the combination $t_{(k+l)N+r} + \theta_{kN,lN+r}/2$.

\section{Extension of Lax hierarchies in Sato theory}
\label{sec:Sato}
\setcounter{equation}{0}
Let $\cal A$ be the algebra of (formal) pseudo-differential operators, the elements of
which are formal series in integer powers of $\partial$ (the derivation with respect
to $x=t_1$). The coefficients of powers of $\partial$ are taken to be
matrices of functions of variables $t_1,t_2, \ldots$, where $t_1 = x$.
The action of powers of the formal inverse $\partial^{-1}$ on functions is given by
\be
     \partial^{-m} f
   = \sum_{j=0}^\infty (-1)^j {m+j-1 \choose j} \, \frac{\partial^j f}{\partial x^j}
     \, \partial^{-m-j}
   \qquad m > 0
\ee
(see \cite{Dick03}, for example).
For $k \in \mathbb{Z}$, the algebra $\cal A$ splits according to
\be
    {\cal A} = {\cal A}_{\geq k} \oplus {\cal A}_{<k}
\ee
where elements of ${\cal A}_{\geq k}$ only contain powers $\geq k$ of $\partial$,
whereas those of ${\cal A}_{< k}$ only contain powers $< k$. But only for $k=0$ and $k=1$
this decomposition satisfies (\ref{decomp1}), i.e. ${\cal A}_{\geq k}$ and ${\cal A}_{<k}$
are Lie subalgebras in these cases.\footnote{In the case of \emph{commutative} coefficients,
one obtains also for $k=2$ a decomposition into Lie subalgebras (see \cite{Blas98},
for example). With \emph{non}commutative coefficients this is no longer so, since in
general the commutator of two vector fields is no longer a vector field:
$[u \, \partial , v \, \partial ]_\ast = ( u \ast v_x - v \ast u_x ) \, \partial
+ [ u,v ]_\ast \, \partial^2$. As a consequence, there is no \emph{non}commutative version
of the Harry-Dym hierarchy (where $L = u_0 \, \partial + u_1 + u_2 \, \partial^{-1}
+ u_3 \, \partial^{-2} + \ldots$), at least not in the framework considered here.}
Choosing
\be
   L = \left\{ \begin{array}{lll}
       \partial + u_2 \, \partial^{-1} + u_3 \, \partial^{-2} + \ldots & \mbox{if } & k=0 \\
       \partial + u_1 + u_2 \, \partial^{-1} + u_3 \, \partial^{-2} + \ldots & \mbox{if } & k=1
       \end{array} \right.
\ee
the equations
\be
    L_{t_n} = [ (L^n)_{\geq k} , L ]_\ast   \qquad  n=1,2, \ldots \label{gncKPhier}
\ee
where $( \; )_{\geq k}$ projects to ${\cal A}_{\geq k}$, determine the \emph{ncKP} ($k=0$)
and \emph{ncmKP} ($k=1$) hierarchies (see \cite{Blas98}, for example, for the commutative
case, and \cite{Kupe00,Hama03,Hama+Toda04,SWW04,Saka04}).
Since for $k =0,1$, ${\cal A}_{\geq k}$ and ${\cal A}_{<k}$ are moreover subalgebras of
$({\cal A},\ast)$, the theorem in section~\ref{sec:ext-MLh} applies.
We thus obtain extensions of the corresponding hierarchies, which we call \emph{xncKP}
(see also \cite{DMH04hier,DMH04ncKP}) and \emph{xncmKP}, respectively.
In the following, $k$ is either 0 or 1. Using the above form of $L$, the identity
$[ (L^n)_{\geq k} , L ]_\ast = - [ (L^n)_{<k} , L ]_\ast$ shows that the right hand
side of (\ref{gncKPhier}) only contains powers of $\partial$ up to the order -1 if $k=0$
and up to order 0 if $k=1$, which is consistent with the left hand side.
\vskip.2cm

Let us introduce the (formal) pseudo-differential operator
\be
    X = w_0 + \sum_{m=1}^\infty w_m \, \partial^{-m}    \label{X}
\ee
with (matrices of) functions $w_m$ depending on the variables $t_n$, $n \in \mathbb{N}$.
Setting $w_0 = 1$ if $k=0$ and assuming $w_0$ to be invertible if $k=1$,
the ncKP and ncmKP hierarchy can be introduced via\footnote{For $k=0$,
this equation apparently appeared first in \cite{Wils80}. See also
\cite{Wils79,Sato82,Mula84} and, in particular, \cite{Kono+Oeve91}, where the
$k=0,1$ \emph{matrix} hierarchies are discussed.}
\be
    X_{t_n} = - (L^n)_{<k} \ast X  \qquad  n=1,2, \ldots  \quad k \in \{ 0,1 \}
    \label{X_tn}
\ee
where\footnote{Equation (\ref{LX}) actually follows from (\ref{X_tn}) with $n=1$:
$- L_{<k} \ast X = X_x = [\partial , X]_\ast = L_{\geq k} \ast X - X \ast \partial$.
Hence $L \ast X = X \ast \partial$.}
\be
     L = X \ast \partial \, X^{-1} \; .  \label{LX}
\ee
The (formal) Baker-Akhiezer function
\be
    \psi = X \ast e^{\sum_{n=1}^\infty t_n \, \lambda^n}
\ee
(with a parameter $\lambda$) then satisfies
\be
     L \ast \psi = \lambda \, \psi \, , \qquad
     \psi_{t_n} = (L^n)_{\geq k} \ast \psi
     \qquad  n=1,2, \ldots  \quad k \in \{ 0,1 \} \; .
\ee
Conversely, (\ref{X_tn}) results as compatibility condition of this linear system.
\vskip.2cm

Equations (\ref{X_tn}) and (\ref{LX}) imply (\ref{gncKPhier}).
According to the discussion in the previous section, the linear system is
consistently extended by (\ref{ext-linsys}) with
\be
    W^{m,n} = \frac{1}{2} \big( (L^n)_{<k} \ast L^m
                  - (L^m)_{<k} \ast L^n \big)_{\geq k} \; .
\ee
In terms of $X$, the deformation equations which extend the nc(m)KP hierarchy
are given by
\be
     X_{\theta_{m,n}}
   = - {1 \over 2} \big( (L^n)_{<k} \ast L^m - (L^m)_{<k} \ast L^n \big)_{<k} \ast X
     \label{X_theta}
\ee
(see \cite{DMH04ncKP} for details in the xncKP case $k=0$).

\subsection{Birkhoff factorization for the xncKP and xncmKP hierarchy}
The nc(m)KP equation (\ref{X_tn}) can be rewritten as\footnote{See \cite{Sato82} for
the case $k=0$. In \cite{OSTT88}, it has been called \emph{Sato equation}.}
\be
    X_{t_n} = - L^n \ast X + (L^n)_{\geq k} \ast X
            = (L^n)_{\geq k} \ast X - X \, \partial^n
\ee
Introducing
\be
    \hat{\xi} = \sum_{n =1}^\infty t_n \,\partial^n
\ee
this can be expressed as follows,
\be
   (X \ast e^{\hat{\xi}})_{t_n} = L^n{}_{\geq k} \ast (X \ast e^{\hat{\xi}}) \; .
     \label{Birk-t}
\ee
In the same way, (\ref{X_theta}) leads to
\bea
      X_{\theta_{m,n}}
  &=& W^{m,n} \ast X - {1 \over 2} \big( (L^n)_{<k} \ast X \, \partial^m
        - (L^m)_{<k} \ast X \, \partial^n \big)   \nonumber \\
  &=& W^{m,n} \ast X + {1 \over 2} ( X_{t_n} \ast \partial^m
      - X_{t_m} \ast \partial^n )
\eea
and thus
\be
  ( X \ast e^{\hat{\xi}} )_{\theta_{m,n}} = W^{m,n} \ast (X \ast e^{\hat{\xi}}) \; .
    \label{Birk-theta}
\ee
\vskip.2cm

\noindent
{\bf Theorem (factorization).}\footnote{See also \cite{Mula84,Taka94,Saka04}.}
The xnc(m)KP equations (\ref{X_tn}) and (\ref{X_theta}) are equivalent to\footnote{Written
in the form $\exp(\hat{\xi}) \ast X_0^{-1} = X^{-1} \ast Y$, this is the
\emph{Birkhoff factorization} (generalized Riemann-Hilbert problem) of
$\exp(\hat{\xi}) \ast X_0^{-1}$. A group element is written in a unique way as the
product of an element of $G_-$ (here the formal group generated by ${\cal A}_{<k}$)
and an element of $G_+$.}
\be
    X \ast e^{\hat{\xi}} \ast X_0^{-1} \in G_+   \label{birk}
\ee
where $X_0$ is the initial value of $X$ at $t := \{ t_n \}_{n=1}^\infty = 0$,
$\theta := \{ \theta_{m,n} \}_{m,n=1}^\infty = 0$, and
$G_+$ is the formal group\footnote{See \cite{Mula84}, in particular.} generated by
the subalgebra ${\cal A}_{\geq k}$ (where $k \in \{0,1\}$).
\vskip.1cm
\noindent
{\em Proof:}
Let us assume that (\ref{X_tn}) and (\ref{X_theta}) hold. We have already shown
that these equations lead to (\ref{Birk-t}) and (\ref{Birk-theta}).
With the help of (\ref{X_tn}) and (\ref{X_theta}), the latter equations imply
\bea
     {\partial \over \partial t_{p_1}} \cdots {\partial \over \partial t_{p_l}}
     {\partial \over \partial \theta_{m_1,n_1}} \cdots
     {\partial \over \partial \theta_{m_r,n_s}}
     ( X \ast e^{\hat{\xi}} )
  = B_{p_1, \ldots, p_l; m_1, n_1, \ldots, m_r, n_s} \ast ( X \ast e^{\hat{\xi}} )
    \nonumber
\eea
where $B_{p_1, \ldots, p_l; m_1, n_1, \ldots, m_r, n_s} \in {\cal A}_{\geq k}$.
Evaluated at $t=0, \theta=0$, these expressions determine (formal) Taylor coefficients
of $X \ast e^{\hat{\xi}}$. Hence $X \ast e^{\hat{\xi}} = Y \ast X_0$ with
\bea
    Y = 1 + \sum_{n=1}^\infty t_n \, B_n\Big|_{t=0,\theta=0}
          + \sum_{m<n} \theta_{m,n} \, B_{m,n}\Big|_{t=0,\theta=0} + \ldots
          \nonumber
\eea
where all higher order terms are also in ${\cal A}_{\geq k}$. Hence $Y$ is an invertible
element of ${\cal A}_{\geq k}$. Furthermore, $Y \in G_+$, since the formal group $G_+$
is defined as the set of formal power series $Y$ (in the variables $t_p, \theta_{m,n}$),
such that $Y-1 \in {\cal A}_{\geq k}$ and $Y-1$ vanishes at
$t=0, \theta=0$.\footnote{Correspondingly, $Z \in G_-$ is defined in the same way,
but with $Z-1 \in {\cal A}_{<k}$. See also \cite{Mula84}.}
Thus (\ref{birk}) holds.

Conversely, let us assume that $X$ satisfies (\ref{birk}) and denote its left hand side
by $Y$, so that
\bea
   X \ast e^{\hat{\xi}} = Y \ast X_0 \; .  \nonumber
\eea
Differentiation with respect to $t_m$ leads to
\bea
     ( X_{t_m} + X \, \partial^m) \ast e^{\hat{\xi}}
  =  ( X \ast e^{\hat{\xi}})_{t_m}
  =  Y_{t_m} \ast X_0
  =  Y_{t_m} \ast Y^{-1} \ast X \ast e^{\hat{\xi}} \nonumber
\eea
and thus
\bea
  X_{t_m} \ast X^{-1} + X \ast \partial^m \ast X^{-1} = Y_{t_m} \ast Y^{-1} \; .
     \nonumber
\eea
The right hand side lies in ${\cal A}_{\geq k}$, since $Y \in G_+$ and
${\cal A}_{\geq k} \ast G_+ \subset {\cal A}_{\geq k}$. Since
$X_{t_m} \ast X^{-1} \in {\cal A}_{<k}$ (see (\ref{X})), by taking the
${\cal A}_{<k}$-part of the last equation we find
\bea
    X_{t_m} = -(X \ast \partial^m \ast X^{-1})_{<k} \ast X   \nonumber
\eea
which is (\ref{X_tn}). In the same way, we obtain
\bea
  \big( X_{\theta_{m,n}} + {1 \over 2} (X_{t_m} \, \partial^n - X_{t_n} \, \partial^m) \big)
    \ast e^{\hat{\xi}}
 &=& (X \ast e^{\hat{\xi}})_{\theta_{m,n}} \nonumber \\
 &=& Y_{\theta_{m,n}} \ast X_0   \nonumber \\
 &=& Y_{\theta_{m,n}} \ast Y^{-1} \ast X \ast e^{\hat{\xi}}   \nonumber
\eea
and therefore
\bea
 \big( X_{\theta_{m,n}} + {1 \over 2} (X_{t_m} \, \partial^n - X_{t_n} \, \partial^m \big)
 \ast X^{-1} = Y_{\theta_{m,n}} \ast Y^{-1} \; .  \nonumber
\eea
Taking the ${\cal A}_{<k}$-part of this equation yields
\bea
    X_{\theta_{m,n}} = - {1 \over 2} ( X_{t_m} \ast \partial^n \ast X^{-1}
      - X_{t_n} \ast \partial^m \ast X^{-1})_{<k} \ast X   \nonumber
\eea
which, by use of (\ref{X_tn}), becomes (\ref{X_theta}). Thus, (\ref{birk})
implies (\ref{X_tn}) and (\ref{X_theta}).
\hfill $\blacksquare$
\vskip.2cm

This result also shows that the deformation equations which extend the Moyal-deformed
hierarchy are in fact a natural property of the latter.

\subsection{xncKP more explicitly}
In this subsection, we take a closer look at the case $k=0$.\footnote{Using
$L_{t_2} = [ (L^2)_{\geq k} , L ]_\ast$, in the $k=0$ case all the coefficients $u_j$
of $L$ with $j>2$ can be expressed (via an $x$-integration) in terms of $u_2$ ($= \phi_x$),
so that the whole hierarchy becomes a set of equations for a single dependent variable.
This does not work in the \emph{non}commutative $k=1$ case. For example, one obtains
$u_{1,t_2} = u_{1,xx} + 2 \, u_1 \ast u_{1,x} + 2 \, u_{2,x} + 2 \, [u_1 , u_2 ]_\ast$,
which cannot be solved for $u_2$ in terms of $u_1$ and its derivatives by an $x$-integration,
since the commutator term also involves $u_2$. See also \cite{Hama03}.}
In terms of
\be
     w_1 = - \phi
\ee
the following formulae for the xncKP hierarchy equations were obtained in \cite{DMH04ncKP}:
\bea
 \phi_{t_m \, t_n} &=& \sigma^{(n)}_{m+1} + (\sigma^{(n)}_{m+1})^\omega
     + \sum_{i=1}^{m-1} \big( \sigma^{(n)}_{m-i} \ast \phi_{t_i}
     + \phi_{t_i} \ast (\sigma^{(n)}_{m-i})^\omega \big) \\
 \phi_{\theta_{m,n}} &=& - {1 \over 2} \Big( \phi_{t_{m+n}}
     + \sigma^{(n)}_{m+1} - (\sigma^{(n)}_{m+1})^\omega  \nonumber \\
 & & + \sum_{i=1}^{m-1} \big( \sigma^{(n)}_{m-i} \ast \phi_{t_i}
     - \phi_{t_i} \ast (\sigma^{(n)}_{m-i})^\omega \big) \Big) \; .
\eea
The coefficients $\sigma^{(n)}_m$ are determined iteratively by
\be
    \sigma^{(n+1)}_m = \sigma^{(1)}_{m,t_n} + \sigma^{(n)}_{m+1} + \sigma^{(1)}_{n+m}
  - \sum_{j=1}^{n-1} \sigma^{(1)}_j \ast \sigma^{(n-j)}_m
  + \sum_{j=1}^{m-1} \sigma^{(1)}_j \ast \sigma^{(n)}_{m-j}
\ee
and
\be
  \sigma^{(1)}_n = p_n(-\tilde{\partial}) \phi
\ee
with $\tilde{\partial} = ( \partial_{t_1}, {1\over 2} \partial_{t_2}, {1 \over 3}
\partial_{t_3}, \ldots )$ and the Schur polynomials
\be
  p_n(t_1,t_2,t_3,\ldots) = \sum_{m_1 + 2 \, m_2 + \ldots + n \, m_n = n \atop m_i \geq 0}
            \frac{1}{m_1! \cdots m_n!} \, t_1^{m_1} \cdots t_n^{m_n} \; .
\ee
Furthermore, ${ }^\omega$ is an involution determined by
\be
  (f_{t_n})^\omega = -(f^\omega)_{t_n} \quad
  (f_{\theta_{m,n}})^\omega = -(f^\omega)_{\theta_{m,n}} \quad
  (f \ast h)^\omega = - h^\omega \ast f^\omega \quad
  \phi^\omega = \phi \, .
\ee
 From these formulae we obtain in particular\footnote{Here $\phi_x{}^2$, for example,
stands for $\phi_x \ast \phi_x$. In our previous papers \cite{DMH04hier,DMH04ncKP}
we wrote $\phi_x{}^{\ast 2}$ instead.}
\bea
  (\phi_{t_3})_x &=& \frac{1}{4} \Big( 3 \, \phi_{yy} + \phi_{xxxx}
       - 6 \, [\phi_x,\phi_y]_\ast + 6 \, (\phi_x{}^2)_x \Big)  \label{phi_t3}  \\
  (\phi_{t_4})_x &=& \frac{1}{3} \Big( 2 \, \phi_{y t_3} + \phi_{xxxy}
       - [\phi_x , 4 \, \phi_{t_3} - \phi_{xxx}]_\ast
       + 3 \, (\{\phi_x , \phi_y\}_\ast)_x \Big) \label{phi_t4} \\
  (\phi_{t_5})_x &=& \frac{1}{216} \Big( 90 \, \phi_{y t_4} + 40 \, \phi_{t_3 t_3}
   + 40 \, \phi_{t_3 xxx} + 45 \, \phi_{xxyy} + \phi_{xxxxxx}  \nonumber \\
  & & + 270 \, [\phi_{t_4} , \phi_x]_\ast
      + 60 \, [\phi_{t_3} , \phi_y + 3 \, \phi_{xx} ]_\ast
      + 120 \, \{\phi_{x t_3} , \phi_x \}_\ast  \nonumber \\
  & & + 45 \, \{\phi_{yy} , \phi_x \}_\ast + 180 \, \{\phi_{xy} , \phi_y \}_\ast
      + 60 \, [ \phi_y , \phi_{xxx} ]_\ast \nonumber \\
  & & + 90 \, [ \phi_x , \phi_{xxy} ]_\ast + 15 \, \{ \phi_{xxxx} , \phi_x \}_\ast \Big)
\eea
and
\bea
 \phi_{\theta_{1,2}} &=& {1 \over 6} (\phi_{t_3} - \phi_{xxx}) - \phi_x{}^2
     \label{phi_theta12} \\
 \phi_{\theta_{1,3}} &=& {1 \over 4} \Big( \phi_{t_4} - \phi_{xxy} - 3 \, \{\phi_x,\phi_y\}_\ast
    - [\phi_x , \phi_{xx}]_\ast \Big)  \label{phi_theta13} \\
 \phi_{\theta_{1,4}}&=& {3 \over 10} \phi_{t_5} - {1 \over 6} \phi_{xxt_3}- {1 \over 8} \phi_{xyy}
  - {1 \over 120} \phi_{xxxxx} - {2 \over 3} \{\phi_x , \phi_{t_3} \}_\ast \nonumber \\
  & & - {1 \over 12} \{\phi_x , \phi_{xxx}\}_\ast
      - {1 \over 4} [\phi_x,\phi_{xy}]_\ast
      + {1 \over 4} [\phi_{xx} , \phi_y]_\ast
      - {1 \over 2} \phi_y{}^2  \label{phi_theta14} \\
 \phi_{\theta_{2,3}} &=&
 {1 \over 10} \phi_{t_5} - {1 \over 8} \phi_{xyy} + {1 \over 40} \phi_{xxxxx}
    - {3 \over 4} \phi_y{}^2 - {1 \over 4} [\phi_x , \phi_{xy}]_\ast  \nonumber \\
  & & + {1 \over 4} \{\phi_x , \phi_{xxx}\}_\ast
      + {1 \over 4} \phi_{xx}{}^2
      + \phi_x{}^3  \label{phi_theta23}
\eea
(see also \cite{DMH04ncKP}). Here $\{ \, , \, \}_\ast$ is the anti-commutator in the
$\ast$-product algebra, and we wrote $y$ instead of $t_2$.
Common $N$-soliton solutions of these equations were obtained in \cite{DMH04ncKP} (see
also \cite{Saka04,Pani01,Wang+Wada03}).

\subsection{$N$-reductions of the xnc(m)KP hierarchy}
Let us consider the $N$-reduction constraint
\be
    (L^N)_{<k} = 0  \qquad N \in \mathbb{N}  \qquad k \in \{ 0,1 \}
\ee
(cf section~\ref{sec:Nred}). In the case $k=0$ it reduces the (non-deformed)
KP hierarchy to the $N$th Gelfand-Dickey (GD) hierarchy \cite{Dick03}. Writing
\be
   (L^N)_{\geq 0} = \partial^N + v_{N-2} \, \partial^{N-2} + v_{N-3} \, \partial^{N-3}
                    + \ldots + v_0
\ee
with (matrices of) functions $v_i$, the $w_j$ can be expressed as
differential polynomials of the $v_i$ (see \cite{Dick03}, for example).
In the case $k=1$, we have
\be
   (L^N)_{\geq 1} = \partial^N + v_{N-1} \, \partial^{N-1}
      + v_{N-2} \, \partial^{N-2} + \ldots + v_1 \, \partial
\ee
instead. The xnc(m)KP equations are then reduced to
\bea
      (L^N)_{t_n}
  &=& [(L^n)_{\geq k}, L^N ]_\ast
      \qquad \forall \, n \in \mathbb{N}: \; n/N \not\in \mathbb{N} \label{Nred-t} \\
      (L^N)_{\theta_{m,n}}
  &=& \frac{1}{2} \big( (L^N)_{t_n} \ast (L^m)_{\geq k}
       - (L^N)_{t_m} \ast (L^n)_{\geq k} \big) + [ W^{m,n}, L^N ]_\ast \qquad
                \label{Nred-theta} \\
  & &   \qquad \qquad \qquad
       \forall \, m,n \in \mathbb{N}: \; m/N, n/N \not\in \mathbb{N} \, .  \nonumber
\eea
\vskip.2cm

For $k=0$, this yields a recipe to obtain the equations of the reduced xncKP hierarchy
easily from those of the xncKP hierarchy presented in the previous subsection in terms
of the potential $\phi$. It amounts to allowing $\phi$ only to depend on
$t_{jN+r}$ and $\theta_{jN+r,lN+s}$, where $j,l = 0,1,2, \ldots$ and $r,s = 1,2,\ldots,N-1$.
Furthermore, derivatives of $\phi$ with respect to $t_{mN}$, $m \in \mathbb{N}$,
have to be dropped, and a derivative with respect to $\theta_{jN,lN+r}$,
where $j \in \mathbb{N}$ and $l \in \mathbb{N} \cup \{ 0 \}$, has to be replaced by
$1/2$ times the derivative with respect to $t_{(j+l)N+r}$.
In particular the last feature provides us with a new way to obtain the flows corresponding
to the $t_n$ with $n>N$ of the $N$th GD hierarchy by extending the Moyal-deformed KP
hierarchy, applying the reduction conditions, and then returning to vanishing deformation:
\bea
 \begin{array}{ccc}
   \mbox{ncKP} & \longrightarrow & \mbox{deformation equations} \\
   {\big\uparrow} & & {\big\downarrow} \\
   \mbox{KP} & \longrightarrow & \mbox{$N$th GD hierarchy}
 \end{array} \nonumber
\eea
Of course, the above method also leads to \emph{extensions} of the Moyal-deformed
GD hierarchies. The cases $N=2$ (xncKdV hierarchy) and $N=3$ (xncBoussinesq hierarchy)
were treated in \cite{DMH04ncKP}. For $N=4$, writing
\be
    L^4 = \partial^4 + u \, \partial^2 + v \, \partial + w
\ee
with (matrices of) functions $u,v,w$, the reduction condition implies
$u = - 4 \, w_{1,x}$ and (\ref{Nred-t}) with $n=2$ leads to
\be
  \left( \begin{array}{c} u \\ v \\ w \end{array} \right)_y
  = \left( \begin{array}{c}
     2 \, ( v_x - u_{xx} ) \\
     2 \, w_x + v_{xx} - 2 \, u_{xxx} - u \, u_x + \frac{1}{2} [ u , v ]_\ast \\
     w_{xx} - \frac{1}{2} ( v \, u_x + u \, u_{xx} + u_{xxxx} ) + \frac{1}{2} [u,w]_\ast
     \end{array} \right)
\ee
(see \cite{Blas98}, for example, for the commutative case).
Elimination of $v,w$, and introduction of the potential $\phi$, using $u = 4 \phi_x$,
leads to the integro-differential equation
\bea
  & & ( D^{-1} \phi_{yy} + \phi_{xxx} + 4 \, \phi_x{}^2 - 2 \, D^{-1} [\phi_x , \phi_y]_\ast )_y
  + 2 \, \{ \phi_{xx} , \phi_y \}_\ast  \nonumber \\
  & & - 2 \, [ \phi_x , D^{-1} \phi_{yy} ]_\ast
      + 4 \, [ \phi_x , D^{-1} [\phi_x,\phi_y]_\ast ]_\ast = 0
\eea
where $D^{-1}$ denotes integration with respect to $x$. Alternatively, we obtain
the last equation more directly from (\ref{phi_t3}) and (\ref{phi_t4})
(where $\phi_{t_4}=0$ as a consequence of the reduction conditions) by elimination
of $\phi_{t_3}$, which involves an integration. Similarly, the deformation equations
of the reduced hierarchy are either obtained from (\ref{Nred-theta}), or more directly
from equations like (\ref{phi_theta12}), (\ref{phi_theta13}) etc:
\bea
   \phi_{\theta_{1,2}} &=& \frac{1}{8} \big( D^{-1} \phi_{yy} - \phi_{xxx}
   - 6 \, \phi_x{}^2 - 2 \, D^{-1} [\phi_x,\phi_y]_\ast \big)  \\
   \phi_{\theta_{1,3}} &=& - {1 \over 4} \big( \phi_{xxy} + 3 \, \{\phi_x,\phi_y\}_\ast
    + [\phi_x , \phi_{xx}]_\ast \big) \; .
\eea

See also \cite{Hama03,Hama+Toda04} for some other reductions of the ncKP and
ncmKP hierarchies.

\section{Extended ncToda lattice hierarchy}
\label{sec:xncToda}
\setcounter{equation}{0}
Let $S$ denote the shift operator $(Sf)_k = f_{k+1}$ acting on functions which
depend on a discrete variable $k \in \mathbb{Z}$. Let $\cal A$ be the algebra of
formal series in $S$, the coefficients of which are matrices of functions of
variables $t_1,t_2, \ldots$. Again, multiplication should obey the Moyal product
rule (\ref{Moyal}) with (\ref{Moyal2}). We have the decomposition
\be
    {\cal A} = {\cal A}_{\geq 0} \oplus {\cal A}_{<0}
\ee
where ${\cal A}_{\geq 0}$ and ${\cal A}_{< 0}$ are the subalgebras of
series with only non-negative, respectively only negative powers of $S$.

With the choice of Lax operator
\be
    L = a \, S^{-1} + b + S    \label{Toda-L}
\ee
the {\em ncToda lattice hierarchy} (see \cite{Blas98}, for example, for the
commutative case, and \cite{Saka04}) is the system of equations\footnote{In
the identity $[(L^n)_{\geq 0},L]_\ast = - [(L^n)_{<0},L]_\ast$,
the left hand side only contains the powers $-1,\ldots,n+1$ of $S$,
whereas on the right side only the powers $-n-1,\ldots,0$ appear.
As a consequence, only the coefficients of $S^{-1}$ and $S^0$ survive.
This is consistent with the left hand side of (\ref{Toda-Laxh}) due to the
form (\ref{Toda-L}) of the Lax operator.}
\be
   L_{t_n} = [(L^n)_{\geq 0},L]_\ast  \qquad \qquad  n = 1,2,3, \ldots
   \label{Toda-Laxh}
\ee
where $(\;)_{\geq 0}$ (respectively $( \; )_{<0}$) projects to ${\cal A}_{\geq 0}$
(respectively ${\cal A}_{<0}$). The coefficient of $S^l$,
$l \in \mathbb{Z}$, is taken by $(\;)_{(l)}$. Since the decomposition satisfies
(\ref{decomp}), it follows that the flows given by (\ref{Toda-Laxh}) commute.
Furthermore, the theorem in section~\ref{sec:ext-MLh} applies, so that an extension
of the hierarchy in the sense of section~\ref{sec:ext-MLh} exists: \emph{xncToda}.

One obtains the following system of equations:
\bea
 a_{t_1} &=& b \ast a - a \ast b_{-1} \nonumber  \\
 b_{t_1} &=& a_{+1} - a  \nonumber \\
 a_{t_2} &=& (a_{+1} + b^2) \ast a - a \ast (a_{-1} + b_{-1}{}^2) \nonumber  \\
 b_{t_2} &=& a_{+1} \ast b + b_{+1} \ast a_{+1} - a \ast b_{-1} - b \ast a  \nonumber \\
 a_{t_3} &=& (a_{+1} \ast b + b \ast a + b\ast a_{+1} + b_{+1} \ast a_{+1}
             + b^3) \ast a  \nonumber \\
         & & - a \ast (a_{-1} \ast b_{-2} + a_{-1} \ast b_{-1} + a \ast b_{-1}
             + b_{-1} \ast a_{-1} + b_{-1}{}^3) \nonumber \\
 b_{t_3} &=& a_{+2} \ast a_{+1} + a_{+1}{}^2 + a_{+1} \ast b^2
             + b_{+1} \ast a_{+1} \ast b + b_{+1}{}^2 \ast a_{+1} \nonumber \\
         & & - a \ast a_{-1} - a^2 - a \ast b_{-1}{}^2
             - b \ast a \ast b_{-1} - b^2 \ast a  \\
     & \vdots & \nonumber
\eea
where $a_{+1}$, $a_{-1}$ abbreviate the actions of $S$, respectively $S^{-1}$, on $a$.

In order to compute the $\theta$-equations, which extend the ncToda hierarchy, we
have to evaluate
\be
 W^{m,n} = {1\over 2} \big( (L^n)_{< 0} \ast L^m
           - (L^m)_{< 0} \ast L^n \big)_{\geq 0} \; .
\ee
In particular, we find
\bea
 W^{1,2} &=& {1\over 2} [b,a]_\ast - {1\over 2} a \, S \\
 W^{1,3} &=& {1\over 2} ([a_{+1} + b^2,a]_\ast + [b,a \ast b_{-1}]_\ast)
             - {1\over 2} \, a \ast (b_{-1} + b + b_{+1}) \, S
             - {1\over 2} a \, S^2 \\
 W^{2,3} &=& {1\over 2} (a \ast[a,b_{-1}]_\ast + [a^2,b]_\ast + [a_{+1} + b^2,a \ast b_{-1}]_\ast
               \nonumber\\
         & & - b \ast [a,b]_\ast \ast b + a_{+1} \ast a \ast b - b \ast a \ast a_{+1})
                \nonumber \\
         & & + {1\over 2} (a^2 + a_{+1} \ast a - (a \ast b_{-1} + b \ast a) \ast( b + b_{+1})
             - b^2 \ast a) \, S  \nonumber \\
         & & - {1\over 2} (a \ast b_{-1} + b \ast a) \, S^2 \; .
\eea
Now (\ref{L_theta_mn}) in particular leads to
\bea
     a_{\theta_{1,2}}
 &=& {1\over 2}(a \ast b_{-1} \ast a - a \ast b \ast a + a_{+1} \ast a \ast b_{-1}
     - b \ast a \ast a_{-1}  \nonumber \\
 & & - b \ast a \ast b_{-1}{}^2 + b^2 \ast a \ast b_{-1}) \\
     b_{\theta_{1,2}}
 &=& {1\over 2} (a^2 - a_{+1}{}^2 - b \ast a \ast b_{-1} + b_{+1} \ast a_{+1} \ast b) \; .
\eea

\section{Remarks}
\label{sec:remarks}
\setcounter{equation}{0}
\noindent
1. Integrable systems where the dependent variables take values in a Moyal algebra (which
arises in particular as a large $N$-limit of models where the dependent variables have
values in a Lie algebra of $N \times N$-matrices) have been studied extensively some time ago,
see \cite{Taka94,Kupe90}, for example. In these models, the Moyal product does \emph{not}
involve the `space-time' coordinates, i.e., the evolution parameters and space coordinates.
In contrast, more recent work concentrated on Moyal-deformation of the space-time coordinates.
In the case of a hierarchy with fields depending on evolution parameters
$t_1,t_2,\ldots$ and corresponding deformation parameters $\theta_{m,n}$, one may
consider sub-hierarchies where indeed the Moyal-product only involves a subset of the $t_n$
which is disjoint from the subset of `evolution times' of the respective sub-hierarchy.
In this sense we are indeed led to Moyal models of the older kind.
\vskip.1cm
\noindent
2. The formalism of section~\ref{sec:ext-MLh} can still be generalized, as becomes
apparent from the work in \cite{DMH04hier}. Let $\cal A$ be the algebra of formal
(integer power) series in a variable $\lambda$ with coefficients multiplied according
to the Moyal product rule (\ref{Moyal}) with (\ref{Moyal2}). We have the decomposition
\be
     {\cal A} = {\cal A}_{\geq 0} \oplus {\cal A}_{<0}
\ee
into the subalgebras of non-negative, respectively negative powers of $\lambda$. Instead
of (\ref{Lax-hier}) we set
\be
    L_{t_n} = [ (\lambda^n \, L)_{\geq 0} , L ]_\ast \; .  \label{Lax-hier2}
\ee
Choosing
\be
    W^{m,n} = \frac{1}{2} \big( (\lambda^n \, L)_{<0} \ast \lambda^m \, L
                - (\lambda^m \, L)_{<0} \ast \lambda^n \, L \big)_{\geq 0}
\ee
it has been shown \cite{DMH04hier} that
\be
    L_{\theta_{m,n}} = [ W^{m,n} , L ]_\ast
    + \frac{1}{2} \big( L_{t_n} \ast (\lambda^m \, L)_{\geq 0}
       - L_{t_m} \ast (\lambda^n \, L)_{\geq 0} )
\ee
extends the hierarchy (\ref{Lax-hier2}) to a larger hierarchy. We thus recover
precisely the corresponding formulae of section~\ref{sec:ext-MLh} with
the formal replacement $\lambda^n L \mapsto L^n$. A concrete example of
the above structure is given by the extended Moyal-deformed AKNS hierarchy
\cite{DMH04hier}.
\vskip.1cm
\noindent
3. A suitable framework for further generalizations can be formulated as follows.
Let ${\cal B}$ be an abelian algebra of linear operators acting on an associative
algebra $({\cal A},\ast)$ such that
\be
   \Theta \circ \mathbf{m}_\ast = \mathbf{m}_\ast \circ \Delta(\Theta)  \qquad
   \forall \, \Theta \in {\cal B}
\ee
with the product map $\mathbf{m}_\ast : {\cal A} \times {\cal A} \rightarrow {\cal A}$,
and a coassociative coproduct of the form
\be
   \Delta( \Theta ) = 1 \otimes \Theta + \Theta \otimes 1 + \Delta'(\Theta) \, , \qquad
   \Delta'(\Theta) = \sum \Theta_{(1)} \otimes \Theta_{(2)}
\ee
with $\Theta_{(1)},\Theta_{(2)} \in {\cal B}$ (using the Sweedler notation, see
\cite{Klim+Schm97}, for example).
Furthermore, let ${\cal L} : {\cal B} \rightarrow {\cal A}$ be a linear map.
The integrability conditions of the linear system
\be
    L \ast \psi = \lambda \, \psi \, , \qquad
    \Theta \, \psi = {\cal L}(\Theta) \ast \psi \quad \forall \Theta \in {\cal B}
\ee
with\footnote{More generally, $\psi$ could be taken to be an element of
a left $\cal A$-module. This would require slight changes of the formalism.}
$L, \psi \in {\cal A}$ are then
\be
   \Theta \, L = [ {\cal L}(\Theta) , L ]_\ast
      - \sum ( \Theta_{(1)} L) \ast {\cal L}(\Theta_{(2)})   \label{Theta-flow}
\ee
and
\bea
  \Theta \, {\cal L}(\Theta') - \Theta \, {\cal L}(\Theta')
  &=& [ {\cal L}(\Theta) , {\cal L}(\Theta') ]_\ast
      - \sum \big( \Theta_{(1)} {\cal L}(\Theta') \big) \ast {\cal L}(\Theta_{(2)})
      \nonumber \\
  & & + \sum \big( \Theta_{(1)}' {\cal L}(\Theta) \big) \ast {\cal L}(\Theta_{(2)}')
      \qquad
\eea
for all $\Theta, \Theta' \in {\cal B}$. The last equation implies that the
$\Theta$-flow given by (\ref{Theta-flow}) commutes with the corresponding flow
associated with $\Theta'$.

In section~\ref{sec:ext-MLh}, where $\ast$ is taken to be the Moyal product, the algebra
$\cal B$ consists of the partial derivative operators
$\partial_{t_p}, \partial_{\theta_{m,n}}$, and the map
$\cal L$ is determined by ${\cal L}(\partial_{t_p}) = (L^p)_+$ and
${\cal L}(\partial_{\theta_{m,n}}) = W^{m,n}$.
According to (\ref{dtheta_star}), the coproduct is given by
\bea
   \Delta(\partial_{t_p}) &=& 1 \otimes \partial_{t_p} + \partial_{t_p} \otimes 1 \\
   \Delta(\partial_{\theta_{m,n}}) &=& 1 \otimes \partial_{\theta_{m,n}}
   + \partial_{\theta_{m,n}} \otimes 1
   + \frac{1}{2} \big( \partial_{t_m} \otimes \partial_{t_n}
   - \partial_{t_n} \otimes \partial_{t_m} \big) \; .
\eea
\vskip.1cm
\noindent
4. One may think of iterating the deformation and extension procedure by introducing
new deformation parameters $\theta_{p,m,n}$ and $\theta_{m,n,r,s}$ with the pairs of
extended hierarchy parameters $t_p, \theta_{m,n}$ and $\theta_{m,n}, \theta_{r,s}$,
respectively, and so forth. Of course the product has then to be changed appropriately
in order to preserve associativity. But this can indeed be done.


\begin{thebibliography}{99}

\bibitem{Blas98}
M. B{\l}aszak: {\it Multi-Hamiltonian Theory of Dynamical Systems}.
Springer, Berlin, 1998.

\bibitem{Dick03}
L.~A. Dickey: {\it Soliton Equations and Hamiltonian Systems}.
World Scientific, Singapore, 2003.

\bibitem{Dorf+Foka92}
I.~Ya. Dorfman and A.~S. Fokas:
Hamiltonian theory over noncommutative rings and integrability in
multidimensions, J. Math. Phys. {\bf 33} (1992) 2504--2514.

\bibitem{Olve98}
P. Olver and V.~V. Sokolov: Integrable evolution equations on associative algebras,
Commun. Math. Phys. {\bf 193} (1998) 245--268.

\bibitem{Kupe00}
B.~A. Kupershmidt: {\it KP or mKP}.
American Math. Soc., Providence, 2000.

\bibitem{Stra01}
I.~A.~B. Strachan:
Deformations of dispersionless KdV hierarchies,
Lett. Math. Phys. {\bf 58} (2001) 129--140.

\bibitem{DMH04hier}
A. Dimakis and F. M\"uller-Hoissen:
Extension of noncommutative soliton hierarchies,
J. Phys. A: Math. Gen. {\bf 37} (2004) 4069--4084.

\bibitem{DMH04ncKP}
A.~Dimakis and F.~M\"uller-Hoissen:
Explorations of the extended ncKP hierarchy,
preprint hep-th/0406112, to appear in J. Phys. A.

\bibitem{Hama03}
M.~Hamanaka:
Commuting flows and conservation laws for noncommutative Lax hierarchies,
preprint hep-th/0311206

\bibitem{Hama+Toda04}
M. Hamanaka and K. Toda:
Towards noncommutative integrable equations,
Proc. Inst. Math. NAS Ukraine {\bf 50} (2004) 404--411.

\bibitem{SWW04}
K.~Shigechi, M.~Wadati, and N.~Wang:
WDVV equation and triple-product relation,
preprint hep-th/0404249.

\bibitem{Saka04}
M.~Sakakibara:
Factorization methods for noncommutative KP and Toda hierarchy,
preprint nlin.SI/0408002.

\bibitem{Wils80}
G. Wilson:
On two constructions of conservation laws for Lax equations,
Quart. J. Math. Oxford {\bf 32} (1981) 491--512.

\bibitem{Wils79}
G. Wilson:
Commuting flows and conservation laws for Lax equations,
Math. Proc. Camb. Phil. Soc. {\bf 86} (1979) 131--143; \\
G. Segal and G. Wilson:
Loop groups and equations of KdV type,
Publ. Math. IHES {\bf 61} (1985) 5--65.

\bibitem{Sato82}
M. Sato and Y. Sato:
Soliton equations as dynamical systems on infinite dimensional
Grassmann manifold,
in {\it Nonlinear Partial Differential Equations in Applied Science},
{\it Lecture Notes in Num. Appl. Anal.}, vol. 5 (Eds. H. Fujita, P.~D. Lax, and G.Strang),
North-Holland, Amsterdam, 1982, pp. 259--271; \\
M. Sato:
The KP hierarchy and infinite-dimensional Grassmann manifolds,
in {\it Theta functions},
{\it Proceedings of Symposia in Pure Mathematics}, vol. 49 (Eds. L. Ehrenpreis and R.C. Gunning),
Amer. Math. Society, Providence, 1989, pp. 51--66.

\bibitem{Mula84}
M. Mulase:
Complete integrability of the Kadomtsev-Petviashvili equation,
Adv. Math. {\bf 54} (1984) 57--66.

\bibitem{Kono+Oeve91}
B.G. Konopelchenko and W. Oevel:
Matrix Sato theory and integrable equations in 2+1 dimensions,
in {\it Nonlinear Evolution Equations and Dynamical Systems}, {\it Proceedings of NEEDS '91}
(Eds. M. Boiti, L. Martina and F. Pempinelli), World Scientific, Singapore, 1992,
pp. 87--96.

\bibitem{OSTT88}
Y. Ohta, J. Satsuma, D. Takahashi, and T. Tokihiro:
An elementary introduction to Sato theory,
Prog. Theor. Phys. Suppl. {\bf 94} (1988) 210--241.

\bibitem{Taka94}
K. Takasaki:
Nonabelian KP hierarchy with Moyal algebraic coefficients,
J. Geom. Phys. {\bf 14} (1994) 332--364.

\bibitem{Pani01}
L.~D. Paniak:
Exact noncommutative KP and KdV multi-solitons,
preprint hep-th/0105185.

\bibitem{Wang+Wada03}
N.~Wang and M.~Wadati:
Noncommutative extension of $\bar{\partial}$-dressing method,
J. Phys. Soc. Japan {\bf 72} (2003) 1366--1373; \\
Exact multi-soliton solutions of noncommutative {KP} equation,
J. Phys. Soc. Japan {\bf 72} (2003) 1881--1888.

\bibitem{Kupe90}
B.~A. Kupershmidt:
Quantizations and integrable systems,
Lett. Math. Phys. {\bf 20} (1990) 19--31.

\bibitem{Klim+Schm97}
A. Klimyk and K. Sch\"udgen:
{\it Quantum Groups and Their Representations}.
Springer, Berlin, 1997.

\end{thebibliography}
\end {document}